\documentclass[preprint,showpacs,preprintnumbers]{revtex4}
\usepackage{graphicx}
\usepackage{dcolumn}
\usepackage{bm}

\begin{document}

\title{Thomson scattering in inhomogeneous plasmas: \\ The Role of the Fluctuation-Dissipation Theorem}
\author{V.V. Belyi}
 
\affiliation{IZMIRAN, Russian Academy of Sciences, Troitsk, Moscow, 108840, Russia}

\date{\today}

\begin{abstract}
A self-consistent kinetic theory of Thomson scattering  of an electromagnetic field by a non-uniform plasma is derived. We draw the readers' attention to the inconsistency in recent results on the Thomson scattering in inhomogeneous plasma, which leads to violation of the Fluctuation-Dissipation Theorem. We show that not only the imaginary part, but also the derivatives of the real part of the dielectric susceptibility determine the amplitude and the width of the Thomson scattering spectral lines. As a result of inhomogeneity, these properties become asymmetric with respect to inversion of the sign of the frequency. A method is proposed for measuring local gradients of the electron density with the aid of Thomson scattering.
\end{abstract}
\pacs{ 52.70.-m, 52.25.Gj, 52.25.Dg, 05.10.Gg}
 \maketitle
\section*{Introduction}
When an electromagnetic wave propagates in a plasma, 
its interaction with  fluctuational oscillations of the plasma may result in scattering of the wave, which  
can be accompanied by a change in its frequency and wave vector. The intensity of scattered waves depends on both the intensity of the incident wave and the level of plasma fluctuations. Since the spectrum of plasma fluctuations exhibits sharp maxima at proper plasma frequencies, the
spectrum of scattered waves will also exhibit sharp maxima at frequencies differing from the frequency of the incident wave by the according frequencies of the plasma
fluctuations. The shift, width and shape of spectral lines carry information on such parameters of the plasma as its density, temperature, mean velocity, ion composition etc.
A method of remote probing of a plasma, termed Thomson scattering, is a powerful plasma diagnostic tool  that is widely employed in measurements of  plasma parameters over a fairly broad range of plasma densities from the ionosphere to strongly coupled  plasma.  In such measurements the plasma must be transparent to the probe electromagnetic radiation. This may be microwave \cite{microwave}, laser \cite{Evans} or X-ray radiation. A comprehensive exposition of the state-of-the-art of X -ray Thomson scattering is presented in Review\cite{Redmer}. 

The differential Thomson scattering cross
section, within an elementary solid angle $d\theta
^{\prime }$ and for a frequency interval $d\omega ^{\prime }$ is described by the
   expression\cite{Dougherty, Akhiezer}:
\begin{equation}
d\Xi =\frac{1}{4\pi }(\frac{e^{2}}{m_{e}c^{2}})^{2}\frac{\omega ^{^{\prime
}2}}{\omega_{0} ^{2}}\sqrt{\frac{\varepsilon (\omega ^{\prime })}{\varepsilon
(\omega_{0} )}}(1+\cos ^{2}\theta )(\delta n_{e}\delta n_{e})_{\mathbf{k\
\omega }}d\theta ^{\prime }d\omega ^{\prime },  \label{a.1}
\end{equation}
 where $\mathbf{k=k^{\prime }-k_{0}}$, ${\omega =\omega ^{\prime }-\omega_{0}; }$  $\mathbf{k_{0}},$ $\mathbf{k^{\prime }},$ ${\omega_{0},}$ ${\omega ^{\prime }}$ are the wave vectors and the frequencies of the incident and scattered electromagnetic fields.
Thus, the problem reduces to finding the spectral characteristics of
electron density fluctuations $(\delta n_{e}\delta n_{e})_{\omega ,\mathbf{k}}$ = $S({\bf k,}\omega )$ - the dynamic electron structure factor (DEFF). The theory of equilibrium and nonequilibrium plasma fluctuations was successfully developed in the second half  of the past century \cite{Klimontovich, Ishimaru, GGK, Dufty}. 
In accordance  with  the Poisson equation, DEFF in a spatially 
homogeneous system is directly linked to the electrostatic field fluctuations. 
In thermodynamic equilibrium, the electrostatic field fluctuations satisfy the famous
Callen-Welton Fluctuation-Dissipation Theorem (FDT) \cite{C-W}, linking their
intensity to the \emph{imaginary} part of the dielectric function $\varepsilon (\omega ,\mathbf{k})$ and to the
temperature T.%
\begin{equation}
(\delta \mathbf{E}\delta \mathbf{E})_{\omega ,\mathbf{k}}=\frac{8\pi \ T%
{\rm Im}\varepsilon (\omega ,\mathbf{k})}{\omega \left\vert \varepsilon
(\omega ,\mathbf{k})\right\vert ^{2}}  \label{a.2}
\end{equation}
Eq. (\ref{a.2}) refers to the steady state, for a space uniform system. 
However, it is not evident that the plasma parameters can be kept \emph{%
constant} in both space and time. Inhomogeneities in space and time of these
quantities will certainly also contribute to the fluctuations. Hence it is  challenging to formulate the generalization of the FDT for inhomogeneous plasma and reformulate accordingly the results for the Thomson scattering. \

An attempt to solve this important problem of describing Thomson scattering
in an inhomogeneous plasma has been made recently in \cite{Kozlowski}. The authors proposed the following {\emph ad hoc} generalization for the quantity $S({\bf k,}\omega )$:
\begin{equation}
S({\bf k,}\omega )=\frac{S({\bf k,}\omega )^{id}}{\left\vert \epsilon ({\bf %
k,}\omega )\right\vert ^{2}},  \label{a.3}
\end{equation}%
"where $S({\bf k,}\omega )^{id}$ is the dynamic structure 
factor for an ideal (noniteracting
gas), and the dielectric (screening) function $\epsilon ({\bf %
k,}\omega )$ in the denominator of Eq. (\ref{a.3}) in a first order 
gradient expansion in microscopic variable is:
\begin{equation}
\epsilon ({\bf k,}\omega )=1+\chi({\bf k,}\omega)=1+(1+i\frac{\partial }{\partial \omega }\frac{%
\partial }{\partial t}-i\frac{\partial }{\partial  {\bf r}}\cdot 
\frac{\partial }{\partial {\bf k}})\chi ^{eq}({\bf k,}\omega),  \label{a.4}
\end{equation}%
 $\chi ^{eq}$ is the susceptibility of the ideal Coulomb plasma. 
The index "eq" labels the susceptibility for a homogeneous system in thermodynamic equilibrium"\cite{Kozlowski}.
Noteworthily, while the authors applied the expansion for  the denominator, the numerator in Eq. (\ref{a.3}) has not been correspondingly expanded. Although, this approximation based on the "physical intuition" reflects some properties of the system, it fails, unfortunately, to satisfy the basic principles. This entails a dramatic inaccuracy of this approach. Namely, this resulted in two 
consequences: Firstly, the obtained result is nonphysical,
 since it contradicts  FDT in the local equilibrium state. 
The FDT for a local equilibrium state was proved by
Balescu \cite{Balescu}. The parameters of a system in a local equilibrium state
 can be changed adiabatically
 on a scale greater than the particle mean free path. Inhomogeneity and  nonstationarity 
of plasma fluctuations are manifested via a non-local dependence upon time \cite{BKW} and coordinates \cite{BKWCPP}. The FDT for a non-local plasma was given in 
our paper \cite{Belyi}. A generalization of the Callen-Welton formula for systems with slowly varying parameters presented in  \cite{Belyi2}. The theory of Langevin equations for slow processes and long time correlations in arbitrary statistical systems has been studied in \cite{Lee}.

 Moreover, authors in theirs numerical simulations \cite{Kozlowski}, took into account gradients in the plasma into dispersion, but not in  the dissipation. Inhomogeneons contribution 
$\mu\frac{\partial }{\partial \mu \mathbf{r}}\cdot 
\frac{\partial }{\partial \mathbf{k}}Im\chi ^{eq}$ is negligibly small with respect
 to the dispersion $Re\chi ^{eq}$  ($Im\chi ^{eq}<<Re\chi ^{eq}$).

 Secondly, the obtained correction due
 to the inhomogeneity in the denominator Eq. (\ref{a.3}) is erroneous for  Langmuir oscillations, especially in the case of small wave numbers $k<k_{D}$, which usually occurs in experiments.
And last but not least: the rigorous kinetic theory predicts
asymmetry of spectral lines in an  inhomogeneous plasma. 
Such asymmetry has been indeed detected in spectroscopic studies of plasma flows in magnetic traps
 \cite{Strun, BelyiStrun}.\

In the present paper, applying the Klimontovich-Langevin approach \cite{Kli} and the time-space multiscale technique, we show that not only the \emph{imaginary} part but also the derivatives of the \emph{real} part of the dielectric susceptibility determine the amplitude and width of spectral lines of the electrostatic field fluctuations and of DEFF, as well. As a result of the inhomogeneity, these properties become asymmetric with respect to  inversion of the sign of the frequency. 
In the kinetic regime the dynamic electron structure
factor is more sensitive to space gradients than the spectral function of
the electrostatic field fluctuations. Note that for simple fluids and gases
a general theory of hydrodynamic fluctuations for nonequilibrium stationary
inhomogeneous states has been developed in \cite{Kirpatrick, BelyiTMF}. 
In particular, it has been found that there exists an asymmetry of the spectrum for 
Brillouin scattering from a fluid in a shear
flow or in a temperature gradient. The situation for the plasma problem we
are considering is, however, quite different.
\
\
\section*{Results}

To treat the problem, a kinetic approach is required, especially when the
wavelength of the fluctuations is larger than the Debye wavelength. To
derive nonlocal expressions for the spectral function of the electrostatic
field fluctuation and for DEFF we adopt the Klimontovich-Langevin
approach to describe kinetic fluctuations \cite{Kli}. A kinetic equation for the
fluctuation $\delta f_{a}$ of the one-particle distribution function (DF)
with respect to the reference state $f_{a}$ is considered. In the general
case the reference state is a nonequilibrium DF which varies in space and
time both on the kinetic scale ( mean free path $l_{ei}$ and interparticle
collision time $\nu _{ei}{}^{-1}$) and, also, on the larger hydrodynamic scales.
These scales are much larger than the characteristic fluctuation time $%
\omega ^{-1}$. In the nonequilibrium case we can, therefore, introduce a
small parameter $\mu =\nu _{ei}/\omega $, which allows us to describe
fluctuations on the basis of a multiple space and time scale analysis.
Obviously, the fluctuations vary on both the \textquotedblright
fast\textquotedblright\ $(\mathbf{r},t)$ and the \textquotedblright
slow\textquotedblright\ $(\mu \mathbf{r},\mu t)$ time and space scales: $%
\delta f_{a}(\mathbf{x,}t)=\delta f_{a}(\mathbf{x},t,\mu t,\mu \mathbf{r})$
and $f_{a}(\mathbf{x,}t)=f_{a}(\mathbf{p},\mu t,\mu \mathbf{r}).$ Here $%
\mathbf{x}$ stands for the phase-space coordinates $(\mathbf{r,p})$. The
Langevin kinetic equation for $\delta f_{a}$ has the form \cite{Kli}
\begin{equation}
\widehat{L}_{a\mathbf{x}t}(\delta f_{a}(\mathbf{x,}t)-\delta f_{a}^{S}(%
\mathbf{x,}t))=-e_{a}\delta \mathbf{E(r},t\mathbf{)\cdot }\frac{\partial
f_{a}(\mathbf{x,}t)}{\partial \mathbf{p}},  \label{b.1}
\end{equation}%
where $e_{a\text{ }}$is the charge of the particle of species $a$, $\delta
\mathbf{E}$ is the electrostatic field fluctuation, and the operator $\widehat{L}%
_{a\mathbf{x}t}$ is defined by
\begin{equation}
\widehat{L}_{a\mathbf{x}t}=\frac{\partial }{\partial t}+\mathbf{v}\cdot
\frac{\partial }{\partial \mathbf{r}}+\widehat{\Gamma }_{a}(\mathbf{x,}t); \\ 
\widehat{\Gamma }_{a}(\mathbf{x,}t)=e_{a}\mathbf{E\cdot }\frac{\partial }{%
\partial \mathbf{p}}-\delta \widehat{I}_{a},
\label{b.2}
\end{equation}
and $\delta \widehat{I}_{a}$ is the linearized collision operator. A model collision operators for plasmas is presented in \cite{BelyiEPL}.

 The Langevin source $\delta f_{a}^{S}$ in Eq. (\ref{b.1}) is determined by the
following equation \cite{Kli}:
\begin{equation}
\widehat{L}_{a\mathbf{x}t}\overline{\delta f_{a}(\mathbf{x,}t)\delta f_{b}(%
\mathbf{x}^{\prime }\mathbf{,}t^{\prime })}^{S}=\delta _{ab}\delta
(t-t^{\prime })\delta (\mathbf{x}-\mathbf{x}^{\prime })f_{b}(\mathbf{x}%
^{\prime }\mathbf{,}t^{\prime }).  \label{b.4}
\end{equation}
The solution of Eq. (\ref{b.1}) has the form
\begin{equation}
\delta f_{a}(\mathbf{x,}t)=\delta f_{a}^{S}(\mathbf{x,}t)-
\sum_{b}\int d\mathbf{x}^{\prime }\int\limits_{-\infty }^{t}dt^{\prime
}G_{ab}(\mathbf{x,}t,\mathbf{x}^{\prime }\mathbf{,}t^{\prime })e_{b}\delta
\mathbf{E(r}^{\prime },t^{\prime }\mathbf{)\cdot }\frac{\partial f_{b}(%
\mathbf{x}^{\prime }\mathbf{,}t^{\prime })}{\partial \mathbf{p}^{\prime }},
\label{b.5}
\end{equation}%
where the Green function $G_{ab}(\mathbf{x,}t,\mathbf{x}^{\prime }\mathbf{,}%
t^{\prime })$ of the operator $\widehat{L}_{a\mathbf{x}t}$ is determined by
\begin{equation}
\widehat{L}_{a\mathbf{x}t}G_{ab}(\mathbf{x,}t,\mathbf{x}^{\prime }\mathbf{,}%
t^{\prime })=\delta _{ab}\delta (\mathbf{x}-\mathbf{x}^{\prime })\delta
(t-t^{\prime })  \label{b.6}
\end{equation}
with the causality condition:
\begin{equation}
G_{ab}(\mathbf{x,}t,\mathbf{x}^{\prime }\mathbf{,}t^{\prime })=0,
\label{b.7}
\end{equation}
when $t<t^{\prime }.$

Thus, $\overline{\delta f_{a}(\mathbf{x,}t)\delta f_{b}(\mathbf{x}^{\prime }%
\mathbf{,}t^{\prime })}^{S}$ and $G_{ab}(\mathbf{x,}t,\mathbf{x}^{\prime }%
\mathbf{,}t^{\prime }) $ are connected by the relation:
\begin{equation}
\overline{\delta f_{a}(\mathbf{x,}t)\delta f_{b}(\mathbf{x}^{\prime }\mathbf{%
,}t^{\prime })}^{S}=G_{ab}(\mathbf{x,}t,\mathbf{x}^{\prime }\mathbf{,}%
t^{\prime })f_{b}(\mathbf{x}^{\prime }\mathbf{,}t^{\prime }),t>t^{\prime }.
\label{b.8}
\end{equation}

For stationary and spatially uniform systems the DF $f_{a}$ and the operator
$\widehat{\Gamma }_{a}$ do not depend on time and space. In this case, the
dependence on time and space of the Green function $G_{ab}(\mathbf{x,}t,%
\mathbf{x}^{\prime }\mathbf{,}t^{\prime })$ is manifested only through the
difference $t-t^{\prime }$ and $\mathbf{r}-\mathbf{r}^{\prime }$. However,
when the DF $f_{a}(\mathbf{p,}\mu \mathbf{r,}\mu t)$ and $\widehat{\Gamma }%
_{a}(\mathbf{p,}\mu \mathbf{r,}\mu t)$ are slowly varying quantities in time
and space, and when nonlocal effects are considered, the time and space
dependence of $G_{ab}(\mathbf{x,}t,\mathbf{x}^{\prime }\mathbf{,}t^{\prime
}) $ is more subtle:
\begin{equation}
G_{ab}(\mathbf{x,}t,\mathbf{x}^{\prime }\mathbf{,}t^{\prime })=G_{ab}(%
\mathbf{p,p}^{\prime },\mathbf{r-r}^{\prime }\mathbf{,}t\mathbf{-}t^{\prime
},\mu \mathbf{r}^{\prime }\mathbf{,}\mu t^{\prime }).  \label{b.9}
\end{equation}%
For the homogeneous case this non-trivial result was obtained for the first
time in our previous work \cite{BKW}. \ This result was extended to
inhomogeneous systems \cite{BKWCPP}. Here we want to stress that the nonlocal
effects appear due to the slow time and space dependencies $\mu \mathbf{r}%
^{\prime }$\textbf{\ }and $\mu t^{\prime }$.

Relationship (\ref{b.9}) is directly linked with the constitutive relation
between the electric displacement and the electric field
\begin{equation}
D_{i}(\mathbf{r},t)=\int d\mathbf{r}^{\prime }\int\limits_{-\infty
}^{t}dt^{\prime }\varepsilon _{ij}(\mathbf{r},\mathbf{r}^{\prime
},t,t^{\prime })E_{j}(\mathbf{r}^{\prime },t^{\prime }).  \label{b.10}
\end{equation}

Previously two kinds of constitutive relations were proposed
phenomenologically for a weakly inhomogeneous and slowly time-varying
medium. Kadomtsev \cite{Kadom} \ formulated the so-called \textit{%
symmetrized} constitutive relation
\begin{equation}
D_{i}(\mathbf{r},t)=\int d\mathbf{r}^{\prime }\int\limits_{-\infty
}^{t}dt^{\prime }\varepsilon _{ij}(\mathbf{r}-\mathbf{r}^{\prime
},t-t^{\prime },\mu \frac{\mathbf{r}+\mathbf{r}^{\prime }}{2},\mu \frac{%
t+t^{\prime }}{2})E_{j}(\mathbf{r}^{\prime },t^{\prime }).  \label{b.11}
\end{equation}
\bigskip Rukhadze and Silin \cite{Silin} proposed a \textit{nonsymmetrized}
constitutive relation%
\begin{equation}
D_{i}(\mathbf{r},t)=\int d\mathbf{r}^{\prime }\int\limits_{-\infty
}^{t}dt^{\prime }\varepsilon _{ij}(\mathbf{r}-\mathbf{r}^{\prime
},t-t^{\prime },\mu \mathbf{r},\mu t)E_{j}(\mathbf{r}^{\prime },t^{\prime }).
\label{b.12}
\end{equation}
Both phenomenological formulations are unsatisfactory. The correct
expression should be
\begin{equation}
D_{i}(\mathbf{r},t)=\int d\mathbf{r}^{\prime }\int\limits_{-\infty
}^{t}dt^{\prime }\varepsilon _{ij}(\mathbf{r}-\mathbf{r}^{\prime
},t-t^{\prime },\mu \mathbf{r}^{\prime },\mu t^{\prime })E_{j}(\mathbf{r}%
^{\prime },t^{\prime }).  \label{b.13}
\end{equation}

In the first order, expansion with respect to $\mu $, Eq. (\ref{b.5}) leads
to
\[
\delta f_{a}(\mathbf{x},t)=\delta f_{a}^{S}(\mathbf{x},t)-\sum_{b}e_{b}\int d%
\mathbf{p}^{\prime }d\mathbf{\rho }\int\limits_{0}^{\infty }d\tau (1-\mu
\tau \frac{\partial }{\partial \mu t}-\mu \mathbf{\rho }\cdot \frac{\partial
}{\partial \mu \mathbf{r}})
\]
\begin{equation}
\ \times G_{ab}(\mathbf{\rho },\tau ,\mathbf{p,p}^{\prime },\mu t,\mu
\mathbf{r})\delta \mathbf{E}(\mathbf{r}-\mathbf{\rho },t-\tau )\cdot \frac{%
\partial f_{b}(\mathbf{p}^{\prime },\mu t,\mu \mathbf{r})}{\partial \mathbf{p%
}^{\prime }},  \label{b.14}
\end{equation}%
with $\mathbf{\rho }=\mathbf{r}-\mathbf{r}^{\prime }$ and $\tau =t-t^{\prime
}.$

Using the Poisson equation and performing the Fourier-Laplace transformation for the fast variables
\begin{equation}
\delta \mathbf{E}(\mathbf{k},\omega )=\int\limits_{0}^{\infty }dt\int d%
\mathbf{r}\delta \mathbf{E}(\mathbf{r},t)\exp (-\Delta t+i\omega t-i\mathbf{k%
}\cdot \mathbf{r}),  \label{b.16}
\end{equation}
we have that the spectral function of the
nonequilibrium electrostatic field fluctuations assumes the form \cite{Belyi}:
\begin{equation}
(\delta \mathbf{E}\delta \mathbf{E})_{\omega ,\mathbf{k}}=\frac{%
32\pi ^{2}}{\left|
\varepsilon (\omega ,\mathbf{k})\right| ^{2}}\sum\limits_{a}e_{a}^{2}Re\int d\mathbf{p}(1+i\mu \frac{\partial }{%
\partial \omega }\frac{\partial }{\partial \mu t}-i\mu \frac{\partial }{\partial
\mathbf{k}}\cdot \frac{\partial }{\partial \mu \mathbf{r}})\frac{1}{k^{2}}%
\widehat{L}_{a\omega \mathbf{k}}^{-1}f_{a}(\mathbf{p},\mu \mathbf{r},\mu t),  \label{b.17}
\end{equation}
where we introduced the effective dielectric function as:
\begin{equation}
\varepsilon (\omega ,\mathbf{k}%
)=1+\sum\limits_{a}\chi _{a}(\omega ,\mathbf{k}); \ 
{\chi }_{a}(\omega ,\mathbf{k})=(1+i\mu \frac{\partial }{\partial
\omega }\frac{\partial }{\partial \mu t}-i\mu \frac{\partial }{\partial \mu \mathbf{r}}%
\cdot \frac{\partial }{\partial \mathbf{k}})\chi^{Neq} _{a}(\omega ,\mathbf{k},\mu t,%
\mu\mathbf{r}),  \label{b.19}
\end{equation}%
where
\begin{equation}
\chi _{a}^{Neq} (\omega ,\mathbf{k},\mu t,\mu \mathbf{r})=-\frac{4\pi ie_{a}^{2}}{k^{2}}%
\int d\mathbf{p}\widehat{L}_{a\omega \mathbf{k}}^{-1}\mathbf{k}\cdot \frac{%
\partial }{\partial \mathbf{p}}f_{a}(\mathbf{p,}\mu t,\mu \mathbf{r})  \label{b.20}
\end{equation}
is the susceptibility for a nonequilibrium plasma.

 If the authors of paper \cite{Kozlowski} deal with the nonequilibrium case, they should use an expression similar to Eq. (\ref{b.17}) in Eq. (\ref{a.3}) as well as
the nonequilibrium susceptibility (\ref{b.20}).\

For the local equilibrium case where the reference state $f^{Leq}_{a}$ is
Maxwellian, we have the identity:
\[
\int d\mathbf{p}(1+i\mu \frac{\partial }{\partial \omega }\frac{\partial }{%
\mu \partial t}-i\mu \frac{\partial }{\partial \mathbf{k}}\mathbf{\cdot }\frac{%
\partial }{\mu \partial \mathbf{r}})\frac{1}{k^{2}}\widehat{L}_{a\omega \mathbf{k%
}}^{-1}f^{Leq}_{a}(\mathbf{p},t,\mathbf{r})
\]
\begin{equation}
=\frac{i}{\omega _{a}}\int d\mathbf{p}f_{a}^{Leq}(\mathbf{p},t,\mathbf{r})-\frac{%
iT_{a}}{4\pi e_{a}^{2}{\omega _{a}}}(1+i\mu \frac{\partial }{\partial
\omega }\frac{\partial }{\partial \mu t}-i\mu \frac{\partial }{\partial \mu \mathbf{r}}%
\cdot \frac{\partial }{\partial \mathbf{k}}){\chi_{a}^{Leq} }(\omega ,\mathbf{k}),
\label{b.21}
\end{equation}
and Eq. (\ref{b.17}) takes the form
\begin{equation}
(\delta \mathbf{E}\delta \mathbf{E})_{\omega ,\mathbf{k}}=\sum_{a}\frac{8\pi
\ T_{a}}{\omega _{a}\left\vert {\varepsilon }\mathbf{(}\omega ,%
\mathbf{k})\right\vert ^{2}\mathbf{\ }}Im{\chi }_{a}(\omega ,%
\mathbf{k}),  \label{b.22}
\end{equation}
where $\omega _{a}=\omega -\mathbf{kV}_{a}$.
For the case of equal temperatures and $\mathbf{V}_{a}=0$  Eq. (\ref{b.22}) satisfies the FDT.
In this case the small parameter $\mu $ is determined on the slower
hydrodynamic scale.
Imaginary part of the 
susceptibility $\chi({\bf k,}\omega)$
 determines the width of the spectral line $(\delta {\bf E}\delta {\bf E})_{\omega ,{\bf k}}$
 near the resonance:
\begin{equation}
\gamma =(Im\chi ^{Leq}+\mu\frac{\partial }{\partial \omega }\frac{\partial }{%
\partial \mu t}Re\chi ^{Leq}-\mu\frac{\partial }{\partial \mu \mathbf{r}}\cdot 
\frac{\partial }{\partial \mathbf{k}}Re\chi ^{Leq})/\frac{\partial }{\partial
\omega }Re\chi ^{Leq}. \label{b.23}
\end{equation}%
In Eq. (\ref{b.23}) there appear additional first-order terms of the small parameter $\mu. $
It is important to note that the $imaginary$ part of the dielectric 
susceptibility is now replaced by the $real$ part, which in the plasma resonance may be greater than 
the $imaginary$ part by the same factor $ \mu^{-1}$. Therefore, the second and third terms in Eq. (\ref{b.23})
in the kinetic regime have an effect comparable to that of the first term. Second-order corrections in the
expansion in $\mu $ only appear in the \emph{imaginary} part
of the susceptibility, and they can be reasonably  neglected. It is therefore
sufficient to retain the first-order corrections to resolve the problem.
 The width of the 
spectral lines Eq. (\ref{b.23}) is affected by new nonlocal terms. 
They are not related to Joule dissipation and appear because of an additional phase
 shift between the induction vector  and the electric field. This phase shift results 
from the finite time needed to set the polarization in the plasma with dispersion \cite{Kravtsov}. 
Such a phase shift in the plasma with space dispersion appears due to the medium 
inhomogeneity.

  For the case when the system parameters are
homogeneous in space but vary in time, the correction to the width of the spectral
lines in Eq. (\ref{b.23}) is
still symmetric with respect to the change in sign of $\omega.$ 
However, when the plasma parameters are space dependent this symmetry is lost. 
The real part of the susceptibility $\chi ^{Leq}({\bf k,}\omega)$ in Eq. (\ref{b.23}) 
is an even function of $\omega. $ This property implies that the contribution of the 
space derivative to the expression for the width
 of the spectral lines is odd function of $\omega.$ Moreover, this term gives rise 
to an anisotropy in $\bf k $ space.

Let us estimate this correction for the plasma mode $(\omega=\omega
_{L})$
\begin{equation}
Re\varepsilon =1-\frac{\omega _{L}^{2}}{\omega ^{2}}(1+3\frac{k^{2}T}{%
m\omega ^{2}}); \ Im\varepsilon =\frac{\nu _{ei}}{\omega }  \label{b.24}
\end{equation}
and
\begin{equation}
{\gamma }=[\nu _{ei}+\frac{2}{n}\frac{\partial n}{\partial t}+6%
\frac{\omega _{L}}{nk_{D}^{2}}\mathbf{k\cdot }\frac{\partial n}{\partial
\mathbf{r}}sgn\omega ]/2.  \label{b.25}
\end{equation}%

 For the spatially homogeneous case there is no difference between the
spectral properties of the longitudinal electric field and of the electron
density, because they are related by the Poisson equation. This statement is no
longer valid when an inhomogeneous plasma is considered. Indeed the
longitudinal electric field is linked to the particle density by the
nonlocal relation:
\begin{equation}
\delta {\bf E}({\bf r},t)=-\frac{\partial }{\partial {\bf r}}%
\sum\limits_{a}e_{a}\int \frac{1}{\left\vert {\bf r-r}^{\prime }\right\vert }%
\delta n_{a}({\bf r}^{\prime },t)d{\bf r}^{\prime }.  \label{b.25a}
\end{equation}
In the same approximation as in Eq. (\ref{b.22}) the expression for 
DEFF for a two-component ($a=e,i$) local equilibrium plasma has 
the form \cite{Belyi}:
\begin{equation}
S^{e}({\bf k,}\omega )=\frac{2n_{e}k^{2}}{\omega_{e} k_{D}^{2}}\left\vert \frac{%
1+\widetilde{\chi }_{i}({\bf k,}\omega )}{\widetilde{\epsilon }({\bf k,}%
\omega )}\right\vert ^{2}{\rm {Im}}\,\widetilde{\chi }_{e}({\bf k,}\omega )+\left\vert 
\frac{\widetilde{\chi }_{e}({\bf k,}\omega )}{\widetilde{\epsilon }({\bf k,}%
\omega )}\right\vert ^{2}\frac{\Theta _{i}}{\Theta _{e}}\frac{2n_{e}k^{2}}{%
\omega_{i} k_{D}^{2}}{\rm {Im}}\,\widetilde{\chi }_{i}({\bf k,}\omega ), \label{b.26}
\end{equation}
where $k_{D}$ is the inverse Debye length,
\begin{equation}
\widetilde{\epsilon }({\bf k,}\omega )=1+\sum\limits_{a}\widetilde{\chi }%
_{a}({\bf k,}\omega ),  \label{b.27}
\end{equation}
\begin{equation}
\widetilde{\chi }_{a}(\omega ,{\bf k})=(1+i\mu\frac{\partial }{\partial \omega }%
\frac{\partial }{\partial \mu t}-i\mu\frac{1}{k^{2}}\frac{\partial }{\partial
\mu r_{i}}k_{j}\frac{\partial }{\partial k_{i}}k_{j})\chi _{a}^{Leq}(\omega ,%
{\bf k},\mu t,\mu {\bf r}).  \label{b.28}
\end{equation}
The inhomogeneous correction in Eq. (\ref{b.28}) $ (\frac{1}{k^{2}}\frac{\partial }{\partial
\mu r_{i}}k_{j}\frac{\partial }{\partial k_{i}}k_{j}Re\chi _{a}^{Leq})$ 
is not the same as in Eq. (\ref{b.23}) $(\frac{\partial }{\partial \mu {\bf r}}\cdot 
\frac{\partial }{\partial {\bf k}}Re\chi^{Leq}).$ 
The origin of this difference is that the Green functions for the electrostatic field 
fluctuations and density particle fluctuations are not the same in an inhomogeneous
situation. As above, we can expand $\widetilde{\varepsilon }(\omega ,\mathbf{k})$ near
the plasma resonance $\omega =\omega _{L}$. Thus, for the Langmuir line,
\begin{equation}
S^{e}({\bf k,}\omega )=\frac{\widetilde{\gamma }}{%
(\omega -\omega _{L}sgn\omega)^{2}+\widetilde{\gamma }^{2}}\frac{2n_{e}k^{2}}{%
\omega k_{D}^{2}\partial Re\varepsilon /\partial \omega }\lfloor _{\omega
=\omega _{L}},  \label{b.29}
\end{equation}
where
\begin{equation}
\widetilde{\gamma }=[Im\varepsilon +\mu\frac{\partial ^{2}Re\varepsilon }{%
\partial \mu t\partial \omega }-\mu\frac{1}{k^{2}}\frac{\partial }{\partial\mu r_{i}}%
k_{j}\frac{\partial }{\partial k_{i}}k_{j}Re\varepsilon ]/\frac{\partial
Re\varepsilon }{\partial \omega }\lfloor _{\omega =\omega _{L}sgn\omega }
\label{b.30}
\end{equation}
is the width of DEFF. An estimate for the plasma mode is then:
\begin{equation}
\widetilde{\gamma }=[\nu _{ei}+\frac{2}{n}\frac{\partial n}{\partial t}+%
\frac{\omega _{L}}{nk^{2}}\mathbf{k\cdot }\frac{\partial n}{\partial \mathbf{%
r}}(1+\frac{9k^{2}}{k_{D}^{2}})sgn\omega ]/2.  \label{b.31}
\end{equation}
From this equation we see that the inhomogeneous correction in Eq. (\ref{b.31}) 
 is greater than the one in Eq. (\ref{b.25}) by the factor $(1+k_{D}^{2}/9k^{2})3/2$ 
 \footnote {The misprint in \cite{Belyi} has been 
detected by T. Beuermann (Univ. Rostock).}. For the same inhomogeneity; i.e., the same gradient of
the density, we plot $S^{e}({\bf k,}\omega )$ 
 together with the $(\delta \mathbf{E}\delta \mathbf{E})_{\omega
,\mathbf{k}} $ as functions of frequency (Fig.1). This
figure shows that the asymmetry of the spectral lines is present both for  
 $S^{e}({\bf k,}\omega )$ and $(\delta \mathbf{E}%
\delta \mathbf{E})_{\omega ,\mathbf{k}}$. However, this effect is more
pronounced in $S^{e}({\bf k,}\omega )$ than in $%
(\delta \mathbf{E}\delta \mathbf{E})_{\omega ,\mathbf{k}}$. 
 Such asymmetry has been indeed detected in inhomogeneous plasma
 \cite{Strun, BelyiStrun}. The asymmetry of lines $S^{e}({\bf k,}\omega )$ 
can be used as a new diagnostic tool 
to measure local gradients in the plasma by Thomson scattering.

The Langmuir line (\ref{b.29}) takes the Lorentz form and the amplitude of the spectral line $A$ is inversely proportional to its width
\begin{equation}
A=\frac{{n_{e}}{k^{2}}}{\widetilde{\gamma }{k_{D}^{2}}}.  
\label{b.32}
\end{equation}
The amplitude of the Langmuir line is seen to be more sensitive to the electron density gradient, than to the line width. Thus, for example, in the case of a density gradient equal to $ {\partial n}/n{\partial r}$=$\nu _{ei}/12v_{T} $ and $k_{D}/k=3$, the red line width decreases by 50 percents, while at the same time the amplitude becomes 2 times larger.
From Eq. (\ref{b.32}) and Eq. (\ref{b.31}) quite a simple formula for calculation of the electron density gradient from the Thomson scattering spectrum follows:
\begin{equation}
\mathbf{k\cdot }\frac{\partial n}{n\partial \mathbf{r}} 
=\frac{\nu _{ei}}{v_{T}}\frac{A^{R}-A^{B}}{A^{R}+A^{B}}\frac{k_{D}}{k_{D}^{2}/k^{2}+9}
=\frac{(\gamma^{R}+\gamma^{B})}{\omega _{L}}\frac{A^{R}-A^{B}}{A^{R}+A^{B}}\frac{k_{D}^{2}}{k_{D}^{2}/k^{2}+9},  
\label{b.33}
\end{equation}
here  $A^{R}$, $A^{B}$ and $\gamma^{R}$, $\gamma^{B}$ are the amplitudes and the half-widths of the red and blue Langmuir satellites, respectively (Fig. 1).    

Thus, intensity and width measurements of the red and blue lines of the spectrum allow to determine the scalar product of the electron density gradient and the scattering vector at a given point. To determine the  vector ${\partial n}/n\partial \mathbf{r}$ it is sufficient to measure the
radiation scattered in three directions simultaneously. 

Similar calculations can also be performed for degenerate high density plasma.

\section*{Conclusion}
 A first-principle kinetic theory of Thomson scattering in a non-uniform plasma is
 constructed, which agrees with the basic FDT and provides quantitatively correct results, that have been confirmed experimentally \cite{Strun, BelyiStrun}. Moreover, our theory provides a novel and unique method of a remote probing and measurement of electron density gradients in plasma; this is based on the demonstrated asymmetry of the Thomson scattering lines. The latter may be important for numerous technological application, e.g. for the tokamak \cite{tokamak}, for the high energy density plasma \cite{Redmer} etc.
Our findings are in a sharp contrast with the results of the recent publication \cite{Kozlowski}, where the suggested ad hoc theory did not agree with the local FDT (which proved to hold valid) and led to quantitatively (and qualitatively) incorrect predictions.


\begin{figure}[hb]
\begin{center}
\includegraphics[width=0.5\textwidth]{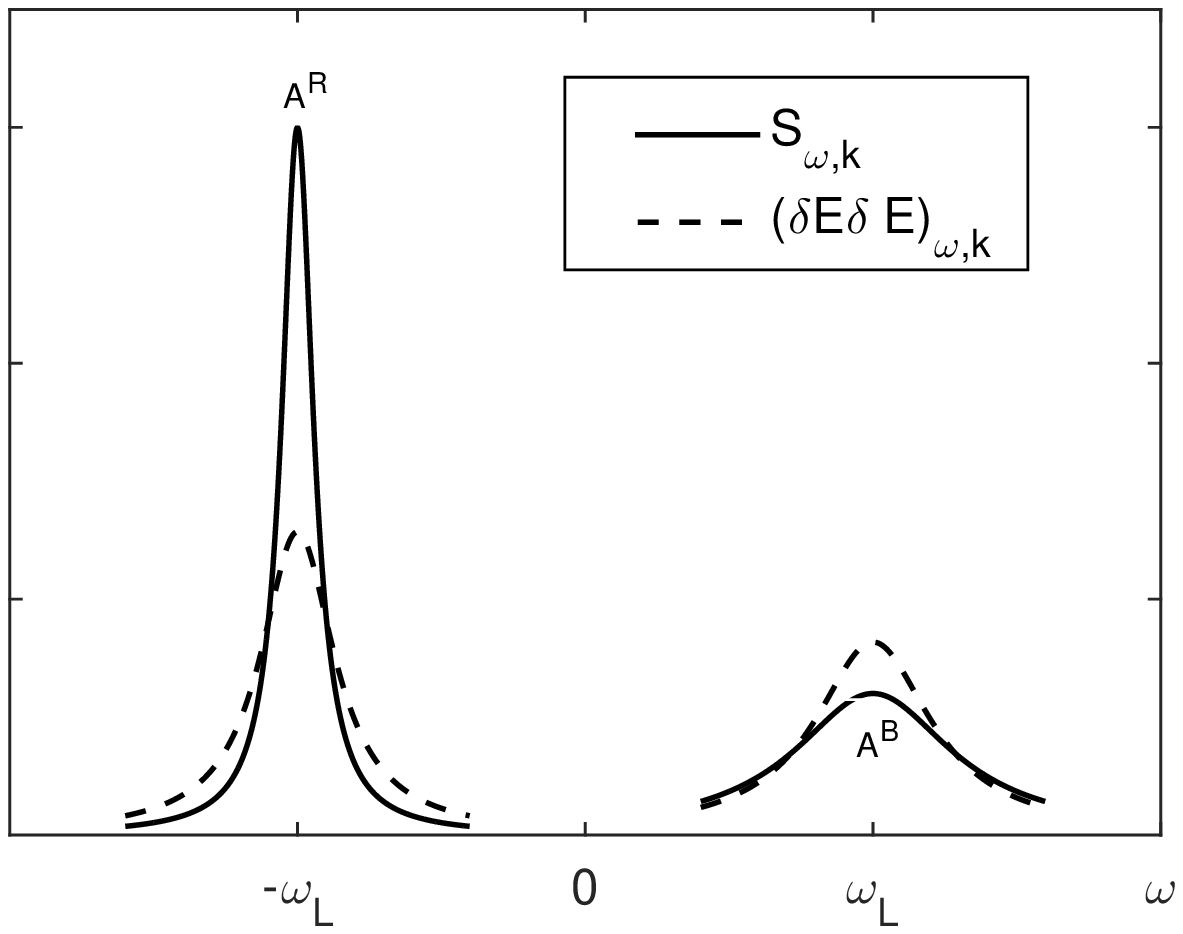}
\caption{The electron structure factor $S^{e}({\bf k,}\omega )$
 ( solid line) and the spectral function of electrostatic field
fluctuations $(\delta {\bf E}\delta {\bf E})_{\omega ,{\bf k}}$ (dashed
line) as a function of frequency. $k_{D}/k=3$; \ ${\bf k\cdot }\frac{\partial n}{n\partial
{\bf r}}=\nu _{ei}k_{D}/27v _{T}$ .
}
\end{center}
\end{figure}

\begin{thebibliography}{}
\bibitem{microwave} V. Heald, C.Wharton, {\it Plasma diagnostics with microwaves} (Wiley, 1965).
\bibitem{Evans} D.Evans, J. Katzenstein, Rep. Progr. Phys. {\bf 32}, 207 (1969).
\bibitem{Redmer} S. Glenzer, R. Redmer, Rev. Mod. Phys.{\bf 81}, 1625 (2009).
\bibitem{Dougherty} J. Dougherty, D. Farley, Proc. Roy. Soc. {\bf A259}, 79 (1960).
\bibitem{Akhiezer} A.I. Akhiezer, I.A. Akhiezer, R.V. Polovin, A.G. Sitenko, K.N. Stepanov, 
{\it Plasma Electrodynamics. V.2. Nonlinear theory and fluctuations} (Pergamon, 1975).
\bibitem{Klimontovich} Yu.L. Klimontovich, {\it The Statistical Theory of Non-Equilibrium Processes in a Plasma}
  (Pergamon, 1967).
\bibitem{Ishimaru} S. Ichimaru, {\it Basic Principles of Plasma Physics}, Ch.7 Fluctuations
(Addison-Wesley, 1973).
\bibitem{GGK}  S.V. Gantsevich, V.L. Gurevich, and R. Katilus, Riv. Nouvo
Cimento. {\bf 2}, 1 (1979).
\bibitem{Dufty} M.C. Marchetti, J.W. Dufty, Physica {\bf 118A}, 205 (1983).
\bibitem{C-W} H.B. Callen, T.A. Welton, Phys. Rev. {\bf 83}, 34 (1951).
\bibitem{Kozlowski} P.M. Kozlowski, B.J.B. Crowley, D.O. Gericke, S.P. Regan, G. Gregori, Sci. Rep. {\bf 6}, 24283 (2016).
\bibitem{Balescu}  R. Balescu, {\it Equlibrium and Nonequlibrium Statistical
Mechanics (Wiley, 1975)}.
\bibitem{BKW}  V.V. Belyi, Yu.A. Kukharenko, and J. Wallenborn, Phys. Rev.
Lett. {\bf 76}, 3554 (1996).
\bibitem{BKWCPP} V.V. Belyi, Yu.A. Kukharenko, J. Wallenborn, Contrib. Plasma Phys. {\bf 42}, 3 (2002). 
\bibitem{Belyi}  V.V. Belyi, Phys. Rev. Lett. {\bf 88}, 255001 (2002).
\bibitem{Belyi2}  V.V. Belyi, Phys. Rev. {\bf E69}, 017104 (2004).
\bibitem{Lee} M.H. Lee, Phys. Rev. {\bf E62}, 1769 (2000).
\bibitem{Strun} V. Strunnikov,{\it Ph.D. thesis}, (Kurchatov Institute, Moscow, 1986).
\bibitem{BelyiStrun} V. Belyi, A. Beschaposhnikov, V. Voronin,V. Strunnikov, Physics of Atomic Nuclei. 
(submitted) (2017).
\bibitem{Kli} Yu. L. Klimontovich, {\it Statistical Physics}
  (Harwood, 1986).
\bibitem{Kirpatrick} T.R. Kirkpatrick, E.G. Cohen, J.R. Dorfman, Phys. Rev. A{\bf 26}, 972 (1982).
\bibitem{BelyiTMF} V. Belyi, Theor. Math. Phys. {\bf 58}, 275 (1984).
\bibitem{BelyiEPL} V. Belyi, Europhys. Lett. {\bf 111}, 40011 (2015).
\bibitem{Kadom} B.B. Kadomtsev, {\it Plasma Turbulence} (Academic Press, 1965).
\bibitem{Silin} A.A. Rukhadze, V.P. Silin, Sov. Phys. Usp. {\bf 7}, 209 (1964).
\bibitem{Kravtsov} M. Bornatici, Yu. Kravtsov, Plasma Phys. Controlled Fusion
 {\bf 42}, 255 (2000).
\bibitem{tokamak} EquipeTFR, Nuclear Fusion {\bf 18}, 647 (1978).
\end{thebibliography}
\end{document}